\documentclass[aps,pra,twocolumn,superscriptaddress]{revtex4}
\usepackage{graphicx}
\usepackage{amssymb}
\usepackage{subfigure}
\usepackage{amsmath}
\usepackage{amsfonts}
\usepackage{bm}
\usepackage{color}
\newcommand{\ket}[1]{\ensuremath{\left|{#1}\right\rangle}}
\newcommand{\bra}[1]{\ensuremath{\left\langle{#1}\right |}}

\newcommand{\oper}[1]{\mathbf{\mathsf{#1}}}

\begin{document}


\title{Ancilla-assisted measurement of photonic spatial correlations and entanglement}

\author{M. Hor-Meyll}
\affiliation{Instituto de F\'{\i}sica, Universidade Federal do Rio de
Janeiro, Caixa Postal 68528, Rio de Janeiro, RJ 21941-972, Brazil}

\author{J. O. de Almeida}
\affiliation{Instituto de F\'{\i}sica, Universidade Federal do Rio de
Janeiro, Caixa Postal 68528, Rio de Janeiro, RJ 21941-972, Brazil}
\author{G. B. Lemos}
\affiliation{Instituto de F\'{\i}sica, Universidade Federal do Rio de
Janeiro, Caixa Postal 68528, Rio de Janeiro, RJ 21941-972, Brazil}
\affiliation{Institute for Quantum Optics and Quantum Information, Boltzmanngasse 3, Vienna A-1090, Austria}
\affiliation{Vienna Center for Quantum Science and Technology (VCQ), Faculty of Physics, University of Vienna, A-1090 Vienna, Austria}
\author{P. H. Souto Ribeiro}
\affiliation{Instituto de F\'{\i}sica, Universidade Federal do Rio de
Janeiro, Caixa Postal 68528, Rio de Janeiro, RJ 21941-972, Brazil}
\author{S. P. Walborn}
\affiliation{Instituto de F\'{\i}sica, Universidade Federal do Rio de
Janeiro, Caixa Postal 68528, Rio de Janeiro, RJ 21941-972, Brazil}
\email[]{swalborn@if.ufrj.br}
\begin{abstract}
We report an experiment in which the moments of spatial coordinates in down-converted photons directly, without having to reconstruct any marginal probability distributions. We use a spatial light modulator to couple the spatial degrees of freedom and the polarization of the fields, which acts as an ancilla system.  Information about the spatial correlations is obtained via measurements on the ancilla qubit.  Among other applications,  this new method provides a more efficient technique to identify continuous variable entanglement.  
\end{abstract}



\maketitle
Continuous variable (CV) entanglement {is a rich platform for fundamental tests of quantum mechanics, and can be a powerful resource for quantum information tasks. Even bipartite CV systems can exhibit high dimensional entanglement\cite{dixon12}, allowing for quantum information protocols that transmit a large amount of information
 \cite{braunstein05}. However, to quantify entanglement  in an infinite dimensional Hilbert space is in general difficult \cite{braunstein05,adesso07}.  In a laboratory setting, even identification of  CV entanglement can be a cumbersome task. CV entanglement witnesses for $d$ degrees of freedom} (DOF) can in general be written in terms of correlation functions involving phase space variables, such as \cite{shchukin05}
\begin{equation}
M_{\vec{\phi},\vec{n}} = \langle x_{\phi_{1}}^{n_{1}}x_{\phi_{2}}^{n_{2}}\cdots x_{\phi_{d}}^{n_{d}} \rangle, 
\label{eq:witness}
\end{equation}
where $\vec{\phi}=(\phi_{1},\dots,\phi_{d})$  specifies the quadrature variable to be measured on each degree of freedom and $\vec{n}=(n_{1},\dots,n_{d})$ are the orders of each moment.   For example, the entanglement witness of Mancini-Giovannetti-Vitali-Tombesi shows that bipartite separable states satisfy \cite{mancini02}  
\begin{equation}
\langle \Delta^2({x}_1\pm {x}_2)\rangle \langle \Delta^2({p}_1\mp {p}_2)\rangle \geq 1,
\label{eq:mancini}
\end{equation}   
where \begin{equation}
\langle \Delta^2({r}_1\pm {r}_2)\rangle =\langle {r}_1^2 \rangle - \langle {r}_1 \rangle^2 + \langle {r}_2^2 \rangle - \langle {r}_2 \rangle^2 \pm 2\langle {r}_1 {r}_2\rangle \mp 2\langle {r}_1\rangle  \langle {r}_2\rangle
\label{eq:mgvt2}
\end{equation}
for variables ${r}={x},{p}$.
Therefore, this and most of the CV separability tests can be cast in terms of the moments of the individual and joint distributions. 
\par
CV entanglement witnesses have been useful to explore the spatial correlations of twin photons produced in parametric down-conversion, which have proven to be a valuable system for investigation of CV entanglement and its applications.  Examples include measurement of entanglement \cite{howell04,dangelo04,tasca08} and Einstein-Podolsky-Rosen-steering correlations \cite{howell04,walborn11a,dixon12,schneeloch13},  applications to quantum cryptography \cite{almeida05}, violation of Bell's inequalities \cite{yarnall07a} and quantum imaging \cite{ribeiro94b,abouraddy01,aspden13}.
 
 \par
 For spatial variables of photons, moments of the sort that appear in Eq. \eqref{eq:witness} can be evaluated by first experimentally determining the marginal probability distributions $P(x_{\phi_1}^{n_1}x_{\phi_2}^{n_2}\cdots x_{\phi_d}^{n_d})$, and then calculating the quantity of interest. 
  For example, to test criteria \eqref{eq:mancini}, one must estimate the distributions $P(x_1,x_2)$ and $P(p_1,p_2)$.  This requires roughly $N^2$ measurements for each distribution, where $N$ is the number of measurement points used for each DOF.   
 {In general, the number of measurements required to reconstruct $P$ to a certain precision is $M \propto N^d$.   Thus, $M$ grows exponentially with the number of DOF $d$.  These considerations are increasingly relevant in the spatial variables domain, as schemes for generation of genuine three-photon spatial entanglement  have been proposed \cite{corona11,avelar13} and four-photon spatial entanglement has been recently observed \cite{torren12}.   
 \par 
  The marginal distributions $P$ contain information about all the possible moments $\vec{n}$ of the relevant variables.  In this regard, evaluating entanglement witnesses using $P$ is not efficient, because in general one would only need to obtain several moments of the type \eqref{eq:witness} in order to evaluate an {entanglement witness $\mathcal{W}$.
 Moreover, using this method, the error  $\delta \mathcal{W}$ typically increases exponentially with $d$, as explicitly shown in Ref. \cite{machado13}. }  These observations hold whether the detection method involves scanning a single-pixel APD or using single-photon sensitive camera \cite{edgar12,aspden13}, since even in the case of a camera, measurement statistics for the complete probability distribution must be obtained. 
\par

  \par
 To overcome these problems, several alternative methods to identify photonic spatial entanglement have been proposed.  One such method is to simply use a coarse-grained detection system with a reduced dimensionality $N$, as in Refs.  \cite{schneeloch13,tasca13}.  However, in this case one accesses only a fraction of the available entanglement.  A promising method uses compressed sensing \cite{howland13} to determine the marginal distribution $P$, which is applicable when $P$ is sparse.  Recently, some of us proposed a method for measuring the moments of spatial variables directly \cite{machado13}, without reconstruction of $P$.  The method provides direct information about a certain moment, and in this sense there is no excess of information obtained.  
 \par
 Here we experimentally demonstrate direct measurement of {spatial moments} that can be used to characterize the spatial entanglement of photon pairs.  Our method is an improved version of the scheme theoretically proposed in Ref. \cite{machado13}.  We directly determine moments for up to $d=4$ degrees of freedom.  
  \paragraph{Ancilla-assisted Measurement.}
\begin{figure}
  \begin{center}
 \includegraphics[width=8cm]{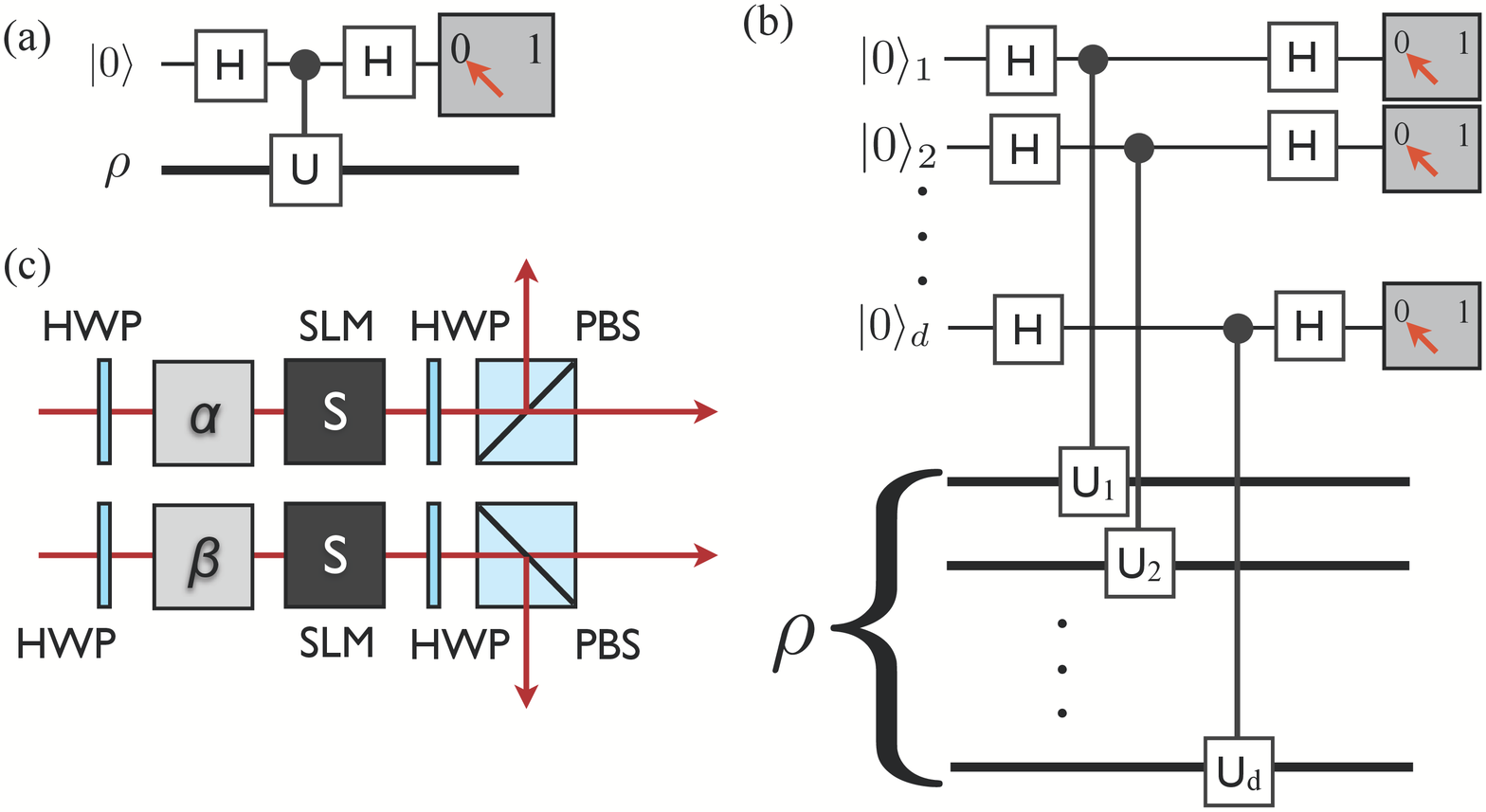}
  \caption{   
(a) Quantum circuit for measurement of the moments of $\rho$. (b) Quantum circuit for measurement of the $d$-mode state $\rho$. c) Conceptual scheme for measurement with spatially-entangled photons.}
\label{fig:setup}
\end{center}
  \end{figure}
We first present our method {to measure the moments of one CV mode with marginal probability distribution $\rho(x_\phi)$,} in terms of the quantum circuit shown in FIG. \ref{fig:setup} a).  The ancilla qubit is prepared in a superposition state $\ket{+}=(\ket{0}+\ket{1})/\sqrt{2}$ using a Hadamard gate, and is then used to control a unitary operator $\oper{U}$.     
After the second Hadamard, the qubit is measured in the computational basis and the CV state is discarded.  
The detection probabilities ($j=0,1$) are \cite{machado13} 
\begin{equation}
Q_j=\mathrm{tr}\{[\oper{U}+(-1)^j\oper{U}^\dagger]\rho[\oper{U}^\dagger+(-1)^j \oper{U}]\}.
\end{equation}
{Defining $\oper{U}=\exp{[i \arccos (\oper{x}^n_\phi)]}$, gives 
\begin{equation}
Q_0-Q_1=\langle {x}^n_\phi\rangle_{\rho}.
\end{equation} 
Hence, by subtracting the measurement outputs in the computational basis, one directly obtains the $n^{\mathrm{th}}$ moment of $\rho(x_\phi)$.  }
\par
{Consider now measurement of the moments of a state $\rho_d$, describing  $d$ CV modes. The state $\rho_d$ is sent through the circuit shown in FIG. \ref{fig:setup} b), together with $d$ auxiliary qubits initially prepared in state $\ket{0}$.}
As shown in Ref. \cite{machado13}, the moments of $\rho_d$ can then be determined directly from the detection probabilities of the qubits by
\begin{equation}
 \langle {x}_{\phi_1}^{n_1}{x}_{\phi_2}^{n_2}\cdots {x}_{\phi_N}^{n_d} \rangle = \sum_{\vec{R}} s^{(\vec{n})}_{\vec{R}}Q_{\vec{R}}, 
 \label{eq:corrfun2}
 \end{equation}
where $R$ is a $d$ bit number encoding the measurement result (ex. $R=7$ if all $d=3$ qubits are detected in state $\ket{1}$).  Here $Q_R$ is the probability of outcome $R$ and $s^{(\vec{n})}_{\vec{R}}=\pm 1$.   The value of $s^{(\vec{n})}_{\vec{R}}$ depends upon the moment to be determined. {Assuming Poissonian count statistics, the uncertainty of the measurement $Q_R$ is}
\begin{equation}
\mathcal{D}^2 = \frac{1}{T}\sum\limits_{R=0}^{2d-1} \left( Q_{\vec{R}} - Q_{\vec{R}}^2\right),
 \label{eq:erroralgo}
 \end{equation}
 where $T$ is the total number of events detected.  Since $\sum_{R=0}^{2d-1} Q_{R} = 1$, this uncertainty is upper-bounded as $\mathcal{D}^2 \leq 1/T$, which does not depend upon the number of DOF.  Therefore, compared to the expected error when complete information of the marginal distributions  $P$ is obtained, this method of directly measuring the moments represents an increase in efficiency that can be exponential with the number of DOF $d$ \cite{machado13}. 
 \par
 Let us now apply this method to the transverse spatial degrees of freedom of entangled photons.  As is customary, we define the near-field variable $x$ and far-field variable $p$ with respect to the output plane of the source \cite{walborn10}. Arbitrary variables $x_{\phi}$ can be accessed using a lens system that implements a fractional Fourier transform \cite{ozaktas01,tasca08}.  In this formalism, $x_{\pi/2} = p$ and $x_{\pi} = -x$. In the spatial DOF of photons, the controlled unitary operator, $\oper{U} = \exp( i \arccos \oper{x}_\phi^n)$, can be implemented with a spatial light modulator (SLM) by exploiting the polarization dependence of this device: the SLM imprints a user defined phase distribution with values between 0 and $\ 2 \pi$ upon the horizontal $H$ polarization component of the wavefront, but not upon the vertically-polarized component $V$. 
 
 FIG. \ref{fig:setup} c) shows a conceptual illustration of the scheme.  We consider that the initial spatial state is $\rho_{12}$, and that the polarization state of the photons is $\ket{HH}_{12}$. The first half-wave plates (HWPs) transform the polarization state to $\ket{++}_{12}$, where $\ket{\pm}=(\ket{H} \pm \ket{V})/\sqrt{2}$. The boxes $\alpha$ and $\beta$ represent lens systems that map either the near-field ($\alpha,\beta=\pi$) or far-field  ($\alpha,\beta=\pi/2$) spatial distributions of the source onto the SLMs.  The action of the SLM on the monochromatic single-photon field can be described as \cite{lemos14} $
 \oper{S} = \ket{H} \bra{H}  \otimes \oper{U} + \ket{V}\bra{V} \otimes \oper{I}$. 
 A second set of HWPs and polarizing beam splitters are used to project onto the polarization states $\ket{\pm}$.   This optical circuit implements the quantum circuit shown in FIG. \ref{fig:setup} b), with the exception that the polarization of each photon controls the $\oper{U}$ operation on {both transverse spatial DOF $x$ and $y$, which correspond to perpendicular directions in the cartesian plane.}   The output probabilities for combinations of $\ket{\pm}$ polarization measurements of photons 1 and 2 are \cite{machado13}:
\begin{align}
P_{++} & = \langle\sin^2 a({x}_{\alpha1},{y}_{\alpha1})\sin^2 b({x}_{\beta2},{y}_{\beta2}) \rangle \\
P_{+-} & = \langle\sin^2 a({x}_{\alpha1},{y}_{\alpha1})\cos^2 b({x}_{\beta2},{y}_{\beta2}) \rangle \\
P_{-+} & = \langle\cos^2 a({x}_{\alpha1},{y}_{\alpha1})\sin^2 b({x}_{\beta2},{y}_{\beta2}) \rangle \\
P_{--} & = \langle\cos^2 a({x}_{\alpha1},{y}_{\alpha1})\cos^2 b({x}_{\beta2},{y}_{\beta2}) \rangle,
\end{align}
{where $a({x}_{\alpha1},{y}_{\alpha1}) = \arccos({x}^{n_1}_{\alpha1},{y}^{m_1}_{\alpha1})$ and $b({x}_{\beta2},{y}_{\beta2}) = \arccos({x}^{n_2}_{\beta2},{y}^{n_2}_{\beta2})$, are the programmed polarization dependent phase shifts imprinted upon the wavefront of each photon by the SLM. We then obtain 
\begin{equation}
\langle {x}_{\alpha1}^{n_1} {y}_{\alpha1}^{m_1} {x}_{\beta2}^{n_2} {y}_{\beta2}^{m_2} \rangle = P_{++} + P_{--} - P_{+-} - P_{-+}.
\label{eq:mom}
\end{equation}
Hence, by programming the SLM to apply the appropriate phases, one can determine any moment of the two-photon spatial distribution directly, without reconstruction of the marginal probability distribution $P(x_{\alpha1},y_{\alpha1},x_{\beta2},y_{\beta2})$.}
  \paragraph{Experiment}
The experimental set-up is shown in FIG. \ref{fig:exp}.
A 325nm He-Cd laser pumps a non-linear crystal (BBO-Beta Barium Borate) producing
spatially entangled photons via spontaneous parametric down-conversion \cite{walborn10}.  The pump beam is focused in the crystal plane using a 1m focal length (FL) lens.  
 We collect pairs of photons with the same wavelength ($\lambda \sim 650$ nm), using high transmittance interference filters with a bandwidth of about
10nm, placed at the entrance of the single photon detectors.  The detectors have large detection apertures, acting as area-integrating ``bucket" detectors, sensitive to the entire spatial extent of the field.    
The down-converted beams are directed to the SLM panel using
mirrors and lenses. These consist of two sets of lenses that can be switched in and out of the
path of the beams with flippable mirror mounts. Depending on the combination of lenses chosen, the image or the optical Fourier
transform  of the source is propagated to the SLM panel, as described above.  The imaging system consists of four confocal lenses (FLs 30cm, 50cm, 20cm, 50cm), and the Fourier transform system of five confocal lenses (FLs 30cm, 30cm, 20cm, 20cm, 50cm).    Both optical systems were chosen so that the down-converted fields covered a substantial region of the SLM. 
\par
The SLM is a reflective phase-only and full-HD-resolution modulator (Holoeye Photonics). The panel is divided in two halves, one for the signal
and the other for the idler photon.  Each side has an area of $960 \times 1080$ pixels and is programmed independently.  After modulation and reflection, the photons are sent
to the detectors through a second optical imaging system and a polarization measurement set-up, consisting of HWPs and PBSs.
This imaging system is composed of two lenses (FLs 70cm and 15cm) in a confocal arrangement, so that a de-magnification factor of $1.5/7\approx 0.21$ is achieved.  This guarantees that all the reflected light falls onto the active area of the detectors.

\begin{figure}
\begin{center}
  \includegraphics[width=8cm]{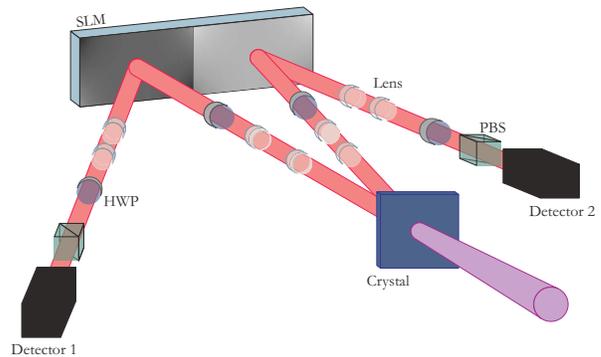}
    \end{center}
  \caption{Experimental setup for ancilla-assisted measurement of spatial correlations.}
 \label{fig:exp}
\end{figure}

\paragraph{Results.}
 \begin{table}
 \caption{\label{tab:1} Experimental results for variance measurements in the near-field variable, in the plane of BBO crystal. }
 \begin{ruledtabular}
 \begin{tabular}{cccc}
Variance  & Direct Method (mm$^2$) & Scanning  (mm$^2$) \\
\hline
$\langle \Delta^2 x_1 \rangle$ & 0.0193 $\pm$ 0.001 &0.0190 $\pm$ 0.0005  \\
$\langle \Delta^2 x_2 \rangle$ & 0.021 $\pm$ 0.002 & 0.0194 $\pm$ 0.0006   \\
$\langle \Delta^2 (x_1 +  x_2) \rangle$ & 0.060 $\pm$ 0.005 &0.067 $\pm$ 0.005  \\
$\langle \Delta^2 (x_1 - x_2) \rangle$ & 0.015 $\pm$ 0.005 & 0.015 $\pm$ 0.005 \\
\end{tabular}
 \end{ruledtabular}
 \end{table}
 \begin{table}
 \caption{\label{tab:2} Experimental results for variance measurements in the far-field variable, in the plane of BBO crystal. }
 \begin{ruledtabular}
 \begin{tabular}{cccc}
Variance  & Direct Method (mm$^{-2}$) & Scanning  (mm$^{-2}$) \\
\hline
$\langle \Delta^2 p_1 \rangle$ & 359 $\pm$ 12 & 386 $\pm$ 19  \\
$\langle \Delta^2 p_2 \rangle$ & 398 $\pm$ 16 & 352 $\pm$ 14   \\
$\langle \Delta^2 (p_1 +  p_2) \rangle$ & 1.64 $\pm$  {34} & 1.77 $\pm 0.32$  \\
$\langle \Delta^2 (p_1 - p_2) \rangle$ & 1419 $\pm$ 34 & 1459 $\pm$ 71 \\

\end{tabular}
 \end{ruledtabular}
 \end{table}
In a first series of measurements, we imaged the
source on the SLM plane for both down-converted fields.  We measured moments $\langle {x}_1\rangle$, $\langle {x}_2\rangle$, $\langle {x}_1^2\rangle$, $\langle {x}_2^2 \rangle$ and $\langle {x}_1{x}_2 \rangle$ by applying the appropriate phase on the SLM.  Here $x$ refers to the transverse vertical direction. A calibration procedure for the SLM is described in the supplementary information \cite{sup}. We then calculated the variances $\langle \Delta^2 {x}_1 \rangle$, $\langle \Delta^2 {x}_2 \rangle$, and $\langle \Delta^2 ({x}_1\pm {x}_2) \rangle$ directly from these measurements using Eq. \eqref{eq:mgvt2}. Results are given in Table \ref{tab:1}. The coordinates at the crystal plane are related to coordinates at the SLM by the demagnification factor $6/25$.  We then tested our results for the direct measurements against the usual method of scanning detectors in the detection plane, {as described in Refs. \cite{howell04,dangelo04,tasca08}}.  To measure the marginal coincidence distributions $C(x_j)$ ($j=1,2$),  the scanning detector was outfitted with a 20$\mu$m slit aperture, while the other detector was left completely open, acting as an area integrating detector.  To measure distributions $C(x_1\pm x_2)$, both detectors were scanned such that $x_1=x_2$ or $x_1=-x_2$.  In all measurements the scanning detectors were displaced in the vertical direction.  These coincidence distributions were used to calculate the variances, also presented in Table \ref{tab:1}.  We obtain a good agreement between both techniques, validating our direct method.     
\par
Switching the optical systems before the SLM to implement an optical Fourier transform of the field distributions in the
source plane onto the SLM panel, we directly measure moments $\langle p_1\rangle$, $\langle p_2\rangle$, $\langle p_1^2\rangle$, $\langle p_2^2 \rangle$, $\langle p_1 p_2\rangle $ of the $p$ (far-field) variables.   The $p$ variable in the crystal plane is obtained from the measurements in the SLM plane by multiplying by the factor $500^{-1}(2\pi/\lambda){\rm  mm} ^{-1}$, where $500$mm is the FL of the Fourier transform lens.   Results are summarized in Table \ref{tab:2}.  Again, these measurements were tested against those obtained from the usual method of scanning detectors in the detection plane.  These results are also shown in Table \ref{tab:2}.  We see good agreement between the two methods.   {We notice that in the case of the momentum variables, a very strong anti-correlation is present leading to a significantly small value for the quantity $\langle\Delta^2( p_1+p_2)\rangle$, which in turn renders an error estimation much larger than the quantity itself.  This is due to the fact the variance $\langle \Delta^2 (p_1 +  p_2) \rangle$ is obtained from the sum of the single-variable variances and covariance using Eq. \eqref{eq:mgvt2}, and due to the strong momentum anti-correlation these nearly cancel.  However, the uncertainty in $\langle \Delta^2 (p_1 +  p_2) \rangle$ is a function of the sum of the individual uncertainties, which are all positive and add up to a larger uncertainty.}
\begin{figure}
\begin{center}
  \includegraphics[width=8cm]{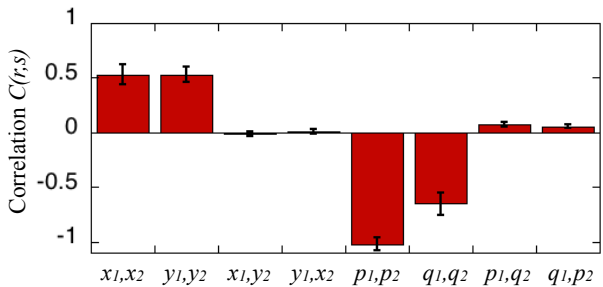}
    \end{center}
  \caption{Correlations $C(r,s)$ for different combinations of near-field $(x,y)$ and far-field $(p,q)$ variables.}
 \label{fig:corr}
\end{figure}
\par
Once the new method was verified, we used it 
to detect entanglement. In order to do so, we measured combinations
of first and second moments in the vertical spatial direction (variables $x$ and $p$), and
also in the horizontal spatial direction (variables $y$ and $q$).
{Moreover, in order to circumvent the problem with a large uncertainty
in the estimation of the sum of momentum, we found that for the direct method, it is advantageous  to}
evaluate the correlations between two variables $r_1$ and $r_2$, defined as \begin{equation}
C(r_1,r_2) = \frac{\langle \Delta r_1 r_2 \rangle} {\sqrt{\langle \Delta^2 r_1 \rangle \langle \Delta^2 r_2\rangle}},
\label{eq:Corr}
\end{equation}
where $\langle \Delta r s \rangle = \langle r s \rangle - \langle r \rangle \langle s \rangle$.  Clearly $-1 \leq C(r_1,r_2) \leq 1$, where $|C|=1$ means perfect correlation. We tested the correlations in the near field variables $x,y$ and the corresponding far-field variables $p,q$.
The results, summarized in FIG. \ref{fig:corr}, show that the photons are correlated in $x_1,x_2$ and $y_1, y_2$, and anti-correlated in $p_1,p_2$ and $q_1,q_2$.  There is less correlation in the far field variables $q$ than in the far-field variables $p$. We attribute this difference to the spatial characteristics of the down-converted photons due to the birefringence of the non-linear crystal \cite{fedorov07}.   We also notice that there is very little cross-axis correlation, since terms like $C(x_1,y_2)$ and $C(p_1,q_2)$, for instance, are quite small.  Using these same correlation functions, we can evaluate an entanglement witnesses directly.  It follows from the Mancini-Giovannetti-Vitali-Tombesi criteria  \eqref{eq:mancini} that all bipartite separable states satisfy
\begin{equation}
|C(r_1,r_2) | + |C(s_1,s_2)| \leq 1, 
\end{equation} 
where $r$ and $s$ are Fourier conjugate variables.  
\par
 We obtained $|C(x_1,x_2)|+|C(p_1,p_2)| =  1.55\pm 0.11$ and  $|C(y_1,y_2)|+|C(q_1,q_2)| =  1.18\pm 0.09$, demonstrating entanglement in both spatial directions, by performing only polarization measurements. 
 \par
\paragraph{Conclusion.}  In conclusion, we have shown that the moments of spatial variables can be obtained directly by imprinting an appropriate phase on the field in an interferometric setup.  In our case, we exploit the polarization dependence of an SLM and observe polarization interference, providing an extremely stable setup.  Using the polarization DOF as an ancilla, we are able to verify entanglement by measuring the polarization and obtaining information about the spatial moments directly, avoiding the need to reconstruct even marginal probability distributions of the entangled photons.  Our results should be increasingly useful for investigations involving more than two photons, as it decreases drastically the number of measurements required.  In addition, we expect that an adaptation of our technique could be useful in exploring time/frequency entanglement, where temporal or spectral measurements are more challenging.  
 \section{Supplementary Information:  Analysis} 
 Let us first consider the case of a single degree of freedom.  The spatial moment $\langle x^n \rangle$ is measured by registering photocounts  $D_{\pm}$ in the $\pm 45^\circ$ polarization direction.  We then calculate $T_{n}$ as
\begin{equation}
T_n = \frac{D_{+} - D_{-}}{D_{+} + D_{-}} = A + B \int I(x_s) x_s^n dx_s,
\label{eq:Tn}
\end{equation}
where $I(x_s)$ is the intensity distribution of the field as a function of the position $x_s$ on the SLM.  The parameters $A$ and $B$ account for experimental errors, due mostly to imperfect operation of the SLM, intensity fluctuations of the pump laser, and imperfect polarization devices. We determine $A$ experimentally by setting the phase of the SLM equal to $a(x)=\pi/2$, which results in $T_{\pi/2}=A$.  At the same time, setting $a(x)=0$ returns $T_0=A+B$.  In this way, we can calculate:
\begin{equation}
\langle x_s^n \rangle_{raw} =  \frac{T_{n}-T_{\frac{\pi}{2}}}{T_{0}-T_{\frac{\pi}{2}}}. 
 \label{eq:xn2}
\end{equation}
This is the spatial moment at the SLM, in units of (pixels/SLM half-width in pixels)$^n$, since $x$ is normalized so that the function $\arccos$ fits on the SLM for $-1 \leq x \leq 1$.  
In order to find the moment in pixels, you must multiply by the total SLM half-width.  The moment in units of pixels$^n$  is 
$\langle x^n \rangle_{pix} = \langle x^n \rangle_{raw} \times (\mathrm{half\, width\, in\, pixels})^n$. 
Converting to millimeters$^n$, we multiply by $(8 \mu$m / pixel)$^n$, to obtain finally $\langle x^n \rangle_{mm}$.  To relate this to the spatial moment at the BBO crystal, we must take into account the lens system used to transfer the field at the BBO to the SLM.  If the BBO to SLM lens system is an imaging system, then we divide by the magnification factor $M^n$.  If it is a lens system used to measure momentum, we can calculate 
\begin{equation}
p_{c} = k \frac{f_2 f_4}{f_1 f_3 f_5} x_{s},
\end{equation}  
where $k=2 \pi / \lambda$. 
\subsection{Higher-dimensional case}
To calculate $\langle x_1^m x_2^n \rangle$ for the two-photon state, we obtain an expression similar to Eq. \eqref{eq:Tn} for each photon. Then, we define:
\begin{equation}
T_{mn} = \frac{D_{++}+D_{--}-D_{+-}-D_{-+}}{D_{++}+D_{--}+D_{+-}+D_{-+}},
\label{eq:T12}
\end{equation}
which, comparing with Eq. \eqref{eq:Tn} gives
\begin{equation}
T_{mn} = A+ M_1\langle x_1^m \rangle + M_2 \langle x_2^n \rangle + N \langle x_1^m x_2^n \rangle.
\label{eq:T12_2}
\end{equation}
The coefficients $A, M_1, M_2$ and $N$ are functions of the different possible errors in the setup, and are determined  experimentally using the following procedure.  We measure constant phases $a(x_1) = b(x_2) = \pi/2$ for each photon.  Then all expectation values are zero, and $T_{\frac{\pi}{2}\frac{\pi}{2}} = A$. We measure $a(x_1) = \pi/2$ and $b(x_2) = \arccos(x_2^n)$, so that all expectation values of $x_1$ are zero, and $T_{\frac{\pi}{2}n} = A + M_2 \langle x_2^n \rangle$. We then measure $a(x_1) = \arccos(x_1^m)$ and $b(x_2) = \pi/2$.  Then  $T_{m\frac{\pi}{2}} = A + M_1 \langle x_1^m \rangle$. Finally, we measure $a(x_1) = \arccos(x_1^m)$ and $b(x_2) = \arccos(x_2^n)$, which gives Eq. \eqref{eq:T12_2} above.   Then we can subtract off the unwanted terms to obtain $N \langle x_1^m x_2^n \rangle$.
We then perform the same steps above for $n=m=0$.  The final calculation is:
\begin{equation}
\langle x_1^m x_2^n \rangle = \frac{T_{nm}-T_{\frac{\pi}{2}n}-T_{m\frac{\pi}{2}}+T_{\frac{\pi}{2}\frac{\pi}{2}}}{T_{00}-T_{\frac{\pi}{2}0}-T_{0\frac{\pi}{2}}+T_{\frac{\pi}{2}\frac{\pi}{2}}}
\end{equation}
The overhead here is that seven measurements are required to determine the moment $\langle x_1^{n_1} x_2^{n_2} \rangle$.   However, we measure moments of the form $\langle x_1^{n_1} y_1^{m_1} x_2^{n_2} y_2^{m_2} \rangle$ with no additional overhead.  The number of total measurements scales as $4N-1$, where $N$ is the number of photons.    
\begin{acknowledgements}
We thank P. Milman for helpful discussions.  We acknowledge financial support from the Brazilian agencies FAPERJ, CNPq,  CAPES and the INCT-Informa\c{c}\~ao Qu\^antica.  SPW acknowledges additional funding from the FET-Open Program of the European Commission under Grant No. 255914 (PHORBITECH).
\end{acknowledgements}
     

\begin{thebibliography}{26}
\expandafter\ifx\csname natexlab\endcsname\relax\def\natexlab#1{#1}\fi
\expandafter\ifx\csname bibnamefont\endcsname\relax
  \def\bibnamefont#1{#1}\fi
\expandafter\ifx\csname bibfnamefont\endcsname\relax
  \def\bibfnamefont#1{#1}\fi
\expandafter\ifx\csname citenamefont\endcsname\relax
  \def\citenamefont#1{#1}\fi
\expandafter\ifx\csname url\endcsname\relax
  \def\url#1{\texttt{#1}}\fi
\expandafter\ifx\csname urlprefix\endcsname\relax\def\urlprefix{URL }\fi
\providecommand{\bibinfo}[2]{#2}
\providecommand{\eprint}[2][]{\url{#2}}

\bibitem[{\citenamefont{Dixon et~al.}(2012)\citenamefont{Dixon, Howland,
  Schneeloch, and Howell}}]{dixon12}
\bibinfo{author}{\bibfnamefont{P.~B.} \bibnamefont{Dixon}},
  \bibinfo{author}{\bibfnamefont{G.~A.} \bibnamefont{Howland}},
  \bibinfo{author}{\bibfnamefont{J.}~\bibnamefont{Schneeloch}},
  \bibnamefont{and} \bibinfo{author}{\bibfnamefont{J.~C.}
  \bibnamefont{Howell}}, \bibinfo{journal}{Phys. Rev. Lett.}
  \textbf{\bibinfo{volume}{108}}, \bibinfo{pages}{143603}
  (\bibinfo{year}{2012}).

\bibitem[{\citenamefont{Braunstein and van Loock}(2005)}]{braunstein05}
\bibinfo{author}{\bibfnamefont{S.~L.} \bibnamefont{Braunstein}}
  \bibnamefont{and} \bibinfo{author}{\bibfnamefont{P.}~\bibnamefont{van
  Loock}}, \bibinfo{journal}{Rev. Mod. Phys.} \textbf{\bibinfo{volume}{77}},
  \bibinfo{pages}{513} (\bibinfo{year}{2005}).

\bibitem[{\citenamefont{Adesso and Illuminati}(2007)}]{adesso07}
\bibinfo{author}{\bibfnamefont{G.}~\bibnamefont{Adesso}} \bibnamefont{and}
  \bibinfo{author}{\bibfnamefont{F.}~\bibnamefont{Illuminati}},
  \bibinfo{journal}{J. Phys. A: Math Theor.} \textbf{\bibinfo{volume}{40}},
  \bibinfo{pages}{7821} (\bibinfo{year}{2007}).

\bibitem[{\citenamefont{Shchukin and Vogel}(2005)}]{shchukin05}
\bibinfo{author}{\bibfnamefont{E.}~\bibnamefont{Shchukin}} \bibnamefont{and}
  \bibinfo{author}{\bibfnamefont{W.}~\bibnamefont{Vogel}},
  \bibinfo{journal}{Phys. Rev. Lett.} \textbf{\bibinfo{volume}{95}},
  \bibinfo{eid}{230502} (\bibinfo{year}{2005}).

\bibitem[{\citenamefont{Mancini et~al.}(2002)\citenamefont{Mancini,
  Giovannetti, Vitali, and Tombesi}}]{mancini02}
\bibinfo{author}{\bibfnamefont{S.}~\bibnamefont{Mancini}},
  \bibinfo{author}{\bibfnamefont{V.}~\bibnamefont{Giovannetti}},
  \bibinfo{author}{\bibfnamefont{D.}~\bibnamefont{Vitali}}, \bibnamefont{and}
  \bibinfo{author}{\bibfnamefont{P.}~\bibnamefont{Tombesi}},
  \bibinfo{journal}{Physical Review Letters} \textbf{\bibinfo{volume}{88}},
  \bibinfo{eid}{120401} (\bibinfo{year}{2002}).

\bibitem[{\citenamefont{Howell et~al.}(2004)\citenamefont{Howell, Bennink,
  Bentley, and Boyd}}]{howell04}
\bibinfo{author}{\bibfnamefont{J.~C.} \bibnamefont{Howell}},
  \bibinfo{author}{\bibfnamefont{R.~S.} \bibnamefont{Bennink}},
  \bibinfo{author}{\bibfnamefont{S.~J.} \bibnamefont{Bentley}},
  \bibnamefont{and} \bibinfo{author}{\bibfnamefont{R.~W.} \bibnamefont{Boyd}},
  \bibinfo{journal}{Phys. Rev. Lett.} \textbf{\bibinfo{volume}{92}},
  \bibinfo{pages}{210403} (\bibinfo{year}{2004}).

\bibitem[{\citenamefont{D'Angelo et~al.}(2004)\citenamefont{D'Angelo, Kim,
  Kulik, and Shih}}]{dangelo04}
\bibinfo{author}{\bibfnamefont{M.}~\bibnamefont{D'Angelo}},
  \bibinfo{author}{\bibfnamefont{Y.-H.} \bibnamefont{Kim}},
  \bibinfo{author}{\bibfnamefont{S.~P.} \bibnamefont{Kulik}}, \bibnamefont{and}
  \bibinfo{author}{\bibfnamefont{Y.}~\bibnamefont{Shih}},
  \bibinfo{journal}{Phys. Rev. Lett.} \textbf{\bibinfo{volume}{92}},
  \bibinfo{pages}{233601} (\bibinfo{year}{2004}).

\bibitem[{\citenamefont{Tasca et~al.}(2008)\citenamefont{Tasca, Walborn,
  Ribeiro, and Toscano}}]{tasca08}
\bibinfo{author}{\bibfnamefont{D.~S.} \bibnamefont{Tasca}},
  \bibinfo{author}{\bibfnamefont{S.~P.} \bibnamefont{Walborn}},
  \bibinfo{author}{\bibfnamefont{P.~H.~S.} \bibnamefont{Ribeiro}},
  \bibnamefont{and} \bibinfo{author}{\bibfnamefont{F.}~\bibnamefont{Toscano}},
  \bibinfo{journal}{Physical Review A} \textbf{\bibinfo{volume}{78}},
  \bibinfo{eid}{010304} (\bibinfo{year}{2008}).

\bibitem[{\citenamefont{Walborn et~al.}(2011)\citenamefont{Walborn, Salles,
  Gomes, Toscano, and Souto~Ribeiro}}]{walborn11a}
\bibinfo{author}{\bibfnamefont{S.~P.} \bibnamefont{Walborn}},
  \bibinfo{author}{\bibfnamefont{A.}~\bibnamefont{Salles}},
  \bibinfo{author}{\bibfnamefont{R.~M.} \bibnamefont{Gomes}},
  \bibinfo{author}{\bibfnamefont{F.}~\bibnamefont{Toscano}}, \bibnamefont{and}
  \bibinfo{author}{\bibfnamefont{P.~H.} \bibnamefont{Souto~Ribeiro}},
  \bibinfo{journal}{Phys. Rev. Lett.} \textbf{\bibinfo{volume}{106}},
  \bibinfo{pages}{130402} (\bibinfo{year}{2011}).

\bibitem[{\citenamefont{Schneeloch et~al.}(2013)\citenamefont{Schneeloch,
  Dixon, Howland, Broadbent, and Howell}}]{schneeloch13}
\bibinfo{author}{\bibfnamefont{J.}~\bibnamefont{Schneeloch}},
  \bibinfo{author}{\bibfnamefont{P.~B.} \bibnamefont{Dixon}},
  \bibinfo{author}{\bibfnamefont{G.~A.} \bibnamefont{Howland}},
  \bibinfo{author}{\bibfnamefont{C.~J.} \bibnamefont{Broadbent}},
  \bibnamefont{and} \bibinfo{author}{\bibfnamefont{J.~C.}
  \bibnamefont{Howell}}, \bibinfo{journal}{Phys. Rev. Lett.}
  \textbf{\bibinfo{volume}{110}}, \bibinfo{pages}{130407}
  (\bibinfo{year}{2013}).

\bibitem[{\citenamefont{Almeida et~al.}(2005)\citenamefont{Almeida, Walborn,
  and Ribeiro}}]{almeida05}
\bibinfo{author}{\bibfnamefont{M.~P.} \bibnamefont{Almeida}},
  \bibinfo{author}{\bibfnamefont{S.~P.} \bibnamefont{Walborn}},
  \bibnamefont{and} \bibinfo{author}{\bibfnamefont{P.~H.~S.}
  \bibnamefont{Ribeiro}}, \bibinfo{journal}{Phys. Rev. A}
  \textbf{\bibinfo{volume}{72}}, \bibinfo{pages}{022313}
  (\bibinfo{year}{2005}).

\bibitem[{\citenamefont{Yarnall et~al.}(2007)\citenamefont{Yarnall, Abouraddy,
  Saleh, and Teich}}]{yarnall07a}
\bibinfo{author}{\bibfnamefont{T.}~\bibnamefont{Yarnall}},
  \bibinfo{author}{\bibfnamefont{A.~F.} \bibnamefont{Abouraddy}},
  \bibinfo{author}{\bibfnamefont{B.~E.~A.} \bibnamefont{Saleh}},
  \bibnamefont{and} \bibinfo{author}{\bibfnamefont{M.~C.} \bibnamefont{Teich}},
  \bibinfo{journal}{Physical Review Letters} \textbf{\bibinfo{volume}{99}},
  \bibinfo{eid}{170408} (\bibinfo{year}{2007}).

\bibitem[{\citenamefont{Ribeiro et~al.}(1994)\citenamefont{Ribeiro, P\'adua,
  da~Silva, and Barbosa}}]{ribeiro94b}
\bibinfo{author}{\bibfnamefont{P.~S.} \bibnamefont{Ribeiro}},
  \bibinfo{author}{\bibfnamefont{S.}~\bibnamefont{P\'adua}},
  \bibinfo{author}{\bibfnamefont{J.~C.~M.} \bibnamefont{da~Silva}},
  \bibnamefont{and} \bibinfo{author}{\bibfnamefont{G.}~\bibnamefont{Barbosa}},
  \bibinfo{journal}{Phys. Rev. A.} \textbf{\bibinfo{volume}{49}},
  \bibinfo{pages}{4176} (\bibinfo{year}{1994}).

\bibitem[{\citenamefont{Abouraddy et~al.}(2001)\citenamefont{Abouraddy, Saleh,
  Sergienko, and Teich}}]{abouraddy01}
\bibinfo{author}{\bibfnamefont{A.~F.} \bibnamefont{Abouraddy}},
  \bibinfo{author}{\bibfnamefont{B.~E.~A.} \bibnamefont{Saleh}},
  \bibinfo{author}{\bibfnamefont{A.~V.} \bibnamefont{Sergienko}},
  \bibnamefont{and} \bibinfo{author}{\bibfnamefont{M.~C.} \bibnamefont{Teich}},
  \bibinfo{journal}{Phys. Rev. Lett.} \textbf{\bibinfo{volume}{87}},
  \bibinfo{pages}{123602} (\bibinfo{year}{2001}).

\bibitem[{\citenamefont{Aspden et~al.}(2013)\citenamefont{Aspden, Tasca, Boyd,
  and Padgett}}]{aspden13}
\bibinfo{author}{\bibfnamefont{R.~S.} \bibnamefont{Aspden}},
  \bibinfo{author}{\bibfnamefont{D.~S.} \bibnamefont{Tasca}},
  \bibinfo{author}{\bibfnamefont{R.~W.} \bibnamefont{Boyd}}, \bibnamefont{and}
  \bibinfo{author}{\bibfnamefont{M.~J.} \bibnamefont{Padgett}},
  \bibinfo{journal}{New Journal of Physics} \textbf{\bibinfo{volume}{15}},
  \bibinfo{pages}{073032} (\bibinfo{year}{2013}).

\bibitem[{\citenamefont{Corona et~al.}(2011)\citenamefont{Corona,
  Garay-Palmett, and U'Ren}}]{corona11}
\bibinfo{author}{\bibfnamefont{M.}~\bibnamefont{Corona}},
  \bibinfo{author}{\bibfnamefont{K.}~\bibnamefont{Garay-Palmett}},
  \bibnamefont{and} \bibinfo{author}{\bibfnamefont{A.~B.} \bibnamefont{U'Ren}},
  \bibinfo{journal}{Opt. Lett.} \textbf{\bibinfo{volume}{36}},
  \bibinfo{pages}{190} (\bibinfo{year}{2011}).

\bibitem[{\citenamefont{Avelar and Walborn}(2013)}]{avelar13}
\bibinfo{author}{\bibfnamefont{A.~T.} \bibnamefont{Avelar}} \bibnamefont{and}
  \bibinfo{author}{\bibfnamefont{S.~P.} \bibnamefont{Walborn}},
  \bibinfo{journal}{Phys. Rev. A} \textbf{\bibinfo{volume}{88}},
  \bibinfo{pages}{032308} (\bibinfo{year}{2013}).

\bibitem[{\citenamefont{van~der Torren et~al.}(2012)\citenamefont{van~der
  Torren, Yorulmaz, Renema, van Exter, and de~Dood}}]{torren12}
\bibinfo{author}{\bibfnamefont{A.~J.~H.} \bibnamefont{van~der Torren}},
  \bibinfo{author}{\bibfnamefont{S.~C.} \bibnamefont{Yorulmaz}},
  \bibinfo{author}{\bibfnamefont{J.~J.} \bibnamefont{Renema}},
  \bibinfo{author}{\bibfnamefont{M.~P.} \bibnamefont{van Exter}},
  \bibnamefont{and} \bibinfo{author}{\bibfnamefont{M.~J.~A.}
  \bibnamefont{de~Dood}}, \bibinfo{journal}{Phys. Rev. A}
  \textbf{\bibinfo{volume}{85}}, \bibinfo{pages}{043837}
  (\bibinfo{year}{2012}).

\bibitem[{\citenamefont{Machado et~al.}(2013)\citenamefont{Machado, Milman, and
  Walborn}}]{machado13}
\bibinfo{author}{\bibfnamefont{S.}~\bibnamefont{Machado}},
  \bibinfo{author}{\bibfnamefont{P.}~\bibnamefont{Milman}}, \bibnamefont{and}
  \bibinfo{author}{\bibfnamefont{S.~P.} \bibnamefont{Walborn}},
  \bibinfo{journal}{Phys. Rev. A} \textbf{\bibinfo{volume}{87}},
  \bibinfo{pages}{053834} (\bibinfo{year}{2013}).

\bibitem[{\citenamefont{Edgar et~al.}(2012)\citenamefont{Edgar, Tasca,
  Izdebski, Warburton, Leach, Agnew, Buller, Boyd, and Padgett}}]{edgar12}
\bibinfo{author}{\bibfnamefont{M.}~\bibnamefont{Edgar}},
  \bibinfo{author}{\bibfnamefont{D.}~\bibnamefont{Tasca}},
  \bibinfo{author}{\bibfnamefont{F.}~\bibnamefont{Izdebski}},
  \bibinfo{author}{\bibfnamefont{R.}~\bibnamefont{Warburton}},
  \bibinfo{author}{\bibfnamefont{J.}~\bibnamefont{Leach}},
  \bibinfo{author}{\bibfnamefont{M.}~\bibnamefont{Agnew}},
  \bibinfo{author}{\bibfnamefont{G.}~\bibnamefont{Buller}},
  \bibinfo{author}{\bibfnamefont{R.}~\bibnamefont{Boyd}}, \bibnamefont{and}
  \bibinfo{author}{\bibfnamefont{M.}~\bibnamefont{Padgett}},
  \bibinfo{journal}{Nature Communications} \textbf{\bibinfo{volume}{3}},
  \bibinfo{pages}{984} (\bibinfo{year}{2012}).

\bibitem[{\citenamefont{Tasca et~al.}(2013)\citenamefont{Tasca, Rudnicki,
  Gomes, Toscano, and Walborn}}]{tasca13}
\bibinfo{author}{\bibfnamefont{D.~S.} \bibnamefont{Tasca}},
  \bibinfo{author}{\bibfnamefont{L.}~\bibnamefont{Rudnicki}},
  \bibinfo{author}{\bibfnamefont{R.~M.} \bibnamefont{Gomes}},
  \bibinfo{author}{\bibfnamefont{F.}~\bibnamefont{Toscano}}, \bibnamefont{and}
  \bibinfo{author}{\bibfnamefont{S.~P.} \bibnamefont{Walborn}},
  \bibinfo{journal}{Phys. Rev. Lett.} \textbf{\bibinfo{volume}{110}},
  \bibinfo{pages}{210502} (\bibinfo{year}{2013}).

\bibitem[{\citenamefont{Howland and Howell}(2013)}]{howland13}
\bibinfo{author}{\bibfnamefont{G.~A.} \bibnamefont{Howland}} \bibnamefont{and}
  \bibinfo{author}{\bibfnamefont{J.~C.} \bibnamefont{Howell}},
  \bibinfo{journal}{Phys. Rev. X} \textbf{\bibinfo{volume}{3}},
  \bibinfo{pages}{011013} (\bibinfo{year}{2013}).

\bibitem[{\citenamefont{Walborn et~al.}(2010)\citenamefont{Walborn, Monken,
  P\'adua, and Ribeiro}}]{walborn10}
\bibinfo{author}{\bibfnamefont{S.~P.} \bibnamefont{Walborn}},
  \bibinfo{author}{\bibfnamefont{C.~H.} \bibnamefont{Monken}},
  \bibinfo{author}{\bibfnamefont{S.}~\bibnamefont{P\'adua}}, \bibnamefont{and}
  \bibinfo{author}{\bibfnamefont{P.~H.~S.} \bibnamefont{Ribeiro}},
  \bibinfo{journal}{Phys. Rep.} \textbf{\bibinfo{volume}{495}},
  \bibinfo{pages}{87} (\bibinfo{year}{2010}).
  
 \bibitem[{\citenamefont{Ozaktas et~al.}(2001)\citenamefont{Ozaktas, Zalevsky,
  and Kutay}}]{ozaktas01}
\bibinfo{author}{\bibfnamefont{H.~M.} \bibnamefont{Ozaktas}},
  \bibinfo{author}{\bibfnamefont{Z.}~\bibnamefont{Zalevsky}}, \bibnamefont{and}
  \bibinfo{author}{\bibfnamefont{M.~A.} \bibnamefont{Kutay}},
  \emph{\bibinfo{title}{The Fractional Fourier Transform: with Applications in
  Optics and Signal Processing}} (\bibinfo{publisher}{John Wiley and Sons Ltd},
  \bibinfo{address}{New York}, \bibinfo{year}{2001}).

\bibitem[{\citenamefont{Lemos et~al.}()\citenamefont{Lemos, Hor-Meyll, Almeida,
  Walborn, and Ribeiro}}]{lemos14}
\bibinfo{author}{\bibfnamefont{G.~B.} \bibnamefont{Lemos}},
  \bibinfo{author}{\bibfnamefont{M.}~\bibnamefont{Hor-Meyll}},
  \bibinfo{author}{\bibfnamefont{J.~O.} \bibnamefont{Almeida}},
  \bibinfo{author}{\bibfnamefont{S.~P.} \bibnamefont{Walborn}},
  \bibnamefont{and} \bibinfo{author}{\bibfnamefont{P.~H.~S.}
  \bibnamefont{Ribeiro}}, \bibinfo{journal}{submitted to PRA}  (2014).

 \bibitem[{\citenamefont{Hor-Meyll et~al.}(2013)\citenamefont{Hor-Meyll, Almeida, Lemos, Ribeiro and Walborn}}]{sup}
  \bibinfo{author}{\bibfnamefont{M.}~\bibnamefont{Hor-Meyll}},
  \bibinfo{author}{\bibfnamefont{J.~O.} \bibnamefont{Almeida}},
  \bibinfo{author}{\bibfnamefont{G.~B.} \bibnamefont{Lemos}},
 \bibnamefont{and} \bibinfo{author}{\bibfnamefont{P.~H.~S.}
  \bibnamefont{Ribeiro}}, 
    \bibinfo{author}{\bibfnamefont{S.~P.} \bibnamefont{Walborn}},
    \bibinfo{journal}{Supplementary Information}.


\bibitem[{\citenamefont{Fedorov et~al.}(2007)\citenamefont{Fedorov, Efremov,
  Volkov, Moreva, Straupe, and Kulik}}]{fedorov07}
\bibinfo{author}{\bibfnamefont{M.~V.} \bibnamefont{Fedorov}},
  \bibinfo{author}{\bibfnamefont{M.~A.} \bibnamefont{Efremov}},
  \bibinfo{author}{\bibfnamefont{P.~A.} \bibnamefont{Volkov}},
  \bibinfo{author}{\bibfnamefont{E.~V.} \bibnamefont{Moreva}},
  \bibinfo{author}{\bibfnamefont{S.~S.} \bibnamefont{Straupe}},
  \bibnamefont{and} \bibinfo{author}{\bibfnamefont{S.~P.} \bibnamefont{Kulik}},
  \bibinfo{journal}{Physical Review Letters} \textbf{\bibinfo{volume}{99}},
  \bibinfo{eid}{063901} (\bibinfo{year}{2007}).

\end{thebibliography}

\end{document}